\documentclass[sigconf]{acmart}

\renewcommand\footnotetextcopyrightpermission[1]{} 
\usepackage{booktabs} 
\usepackage{algorithm, algorithmicx, algpseudocode}
\usepackage{subcaption}
\usepackage{threeparttable}
\usepackage{listings}
\setcopyright{rightsretained}

\acmConference[iPres2018]{ACM Woodstock conference}{September 2018}{Boston and Cambridge, MA USA}
\acmYear{2018}
\copyrightyear{2018}

\hypersetup{draft}

\begin{document}
\title{The Off-Topic Memento Toolkit}

\author{Shawn M. Jones}
\affiliation{%
  \institution{Old Dominion University}
  \city{Norfolk}
  \state{Virginia}
  \postcode{23529}
}
\email{sjone@cs.odu.edu}

\author{Michele C. Weigle}
\affiliation{%
  \institution{Old Dominion University}
  \city{Norfolk}
  \state{Virginia}
  \postcode{23529}}
 \email{mweigle@cs.odu.edu}
 
\author{Michael L. Nelson}
\affiliation{%
  \institution{Old Dominion University}
  \city{Norfolk}
  \state{Virginia}
  \postcode{23529}}
\email{mln@cs.odu.edu}

\renewcommand{\shortauthors}{S. Jones et al.}

\begin{abstract}
Web archive collections are created with a particular purpose in mind. A curator selects seeds, or original resources, which are then captured by an archiving system and stored as archived web pages, or mementos. The systems that build web archive collections are often configured to revisit the same original resource multiple times. This is incredibly useful for understanding an unfolding news story or the evolution of an organization. Unfortunately, over time, some of these original resources can go off-topic and no longer suit the purpose for which the collection was originally created. They can go off-topic due to web site redesigns, changes in domain ownership, financial issues, hacking, technical problems, or because their content has moved on from the original topic. Even though they are off-topic, the archiving system will still capture them, thus it becomes imperative to anyone performing research on these collections to identify these off-topic mementos. Hence, we present the Off-Topic Memento Toolkit, which allows users to detect off-topic mementos within web archive collections. The mementos identified by this toolkit can then be separately removed from a collection or merely excluded from downstream analysis. The following similarity measures are available: byte count, word count, cosine similarity, Jaccard distance, S{\o}rensen-Dice distance, Simhash using raw text content, Simhash using term frequency, and Latent Semantic Indexing via the gensim library. We document the implementation of each of these similarity measures. We possess a gold standard dataset generated by manual analysis, which contains both off-topic and on-topic mementos. Using this gold standard dataset, we establish a default threshold corresponding to the best $F_1$ score for each measure. We also provide an overview of potential future directions that the toolkit may take.
\end{abstract}


%
%

\ccsdesc[500]{Applied computing~Digital libraries and archives}
\ccsdesc[300]{Information systems~World Wide Web}

\keywords{topic drift, Archive-It, web archive, similarity}

\maketitle

\section{Introduction}

Public web archives are where web pages are preserved and accessible for curiosity, research, and evidentiary purposes \cite{Milligan2012, MEET:MEET14504801096}. Some researchers go so far as to curate their own collections of archived web pages, or \textbf{mementos}. These curators will select \textbf{seeds}, or \textbf{original resources}, and create their own mementos from these seeds using variety of web archiving platforms, one of which is the Archive-It service \cite{mcclure2006} offered by the Internet Archive. Collections are created for some purpose and these seeds are chosen to support the collection's topic. In order to understand the history of an event or an organization, curators will often configure the platform to capture the same original resource multiple times, thus producing many mementos per seed. Many collections at Archive-It are like \emph{Japan Earthquake}, with 81,014 seeds resulting in 486,227 mementos. Researchers examining these collections want to optimize the amount of time spent evaluating mementos, and the sheer quantity of mementos to evaluate makes it imperative that they not spend time on mementos with low information value.

\begin{figure*}[t]
\centering
\includegraphics[width=0.4\textwidth]{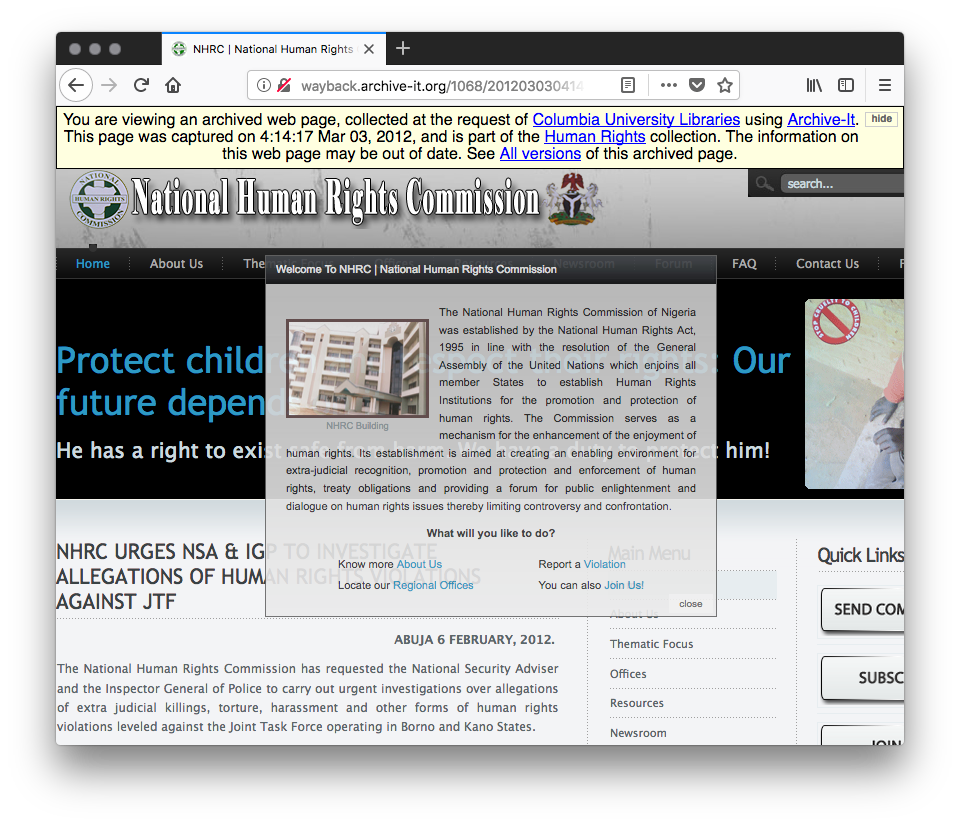}
\includegraphics[width=0.4\textwidth]{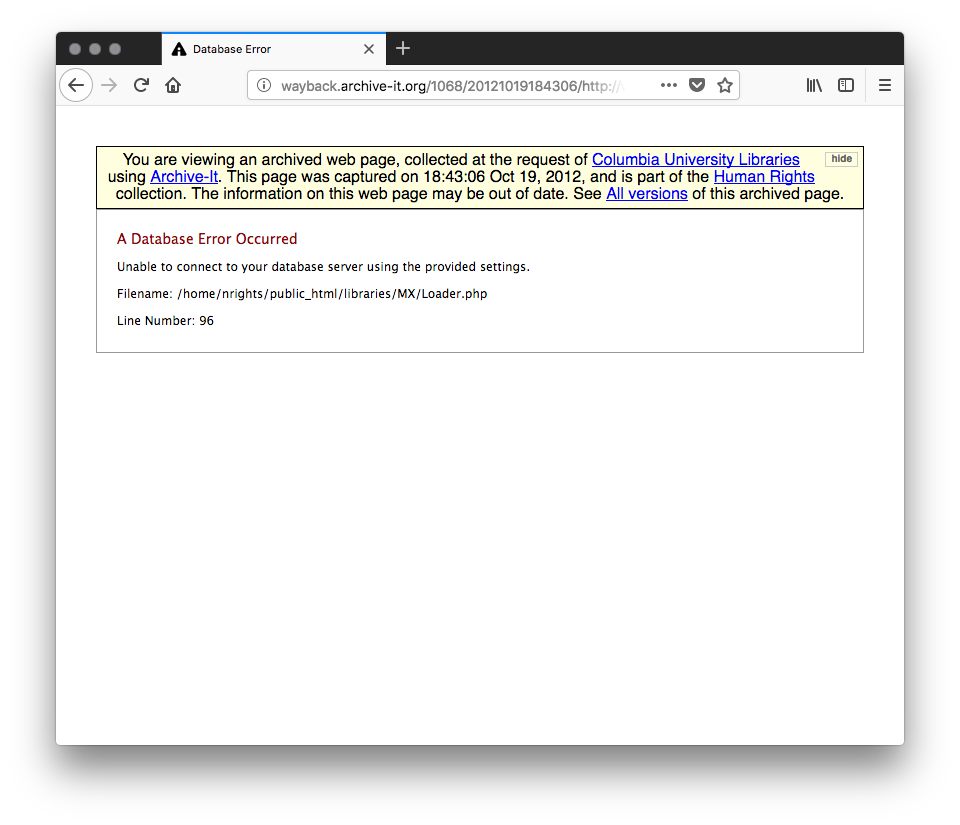}
\caption{The seed \url{http://www.nigeriarights.gov.ng/} preserved in Archive-It's \emph{Human Rights} collection (left) goes off topic due to technical issues (right).}
\label{fig:tech_issues}
\end{figure*}

\begin{figure*}[t]
\includegraphics[width=0.4\textwidth]{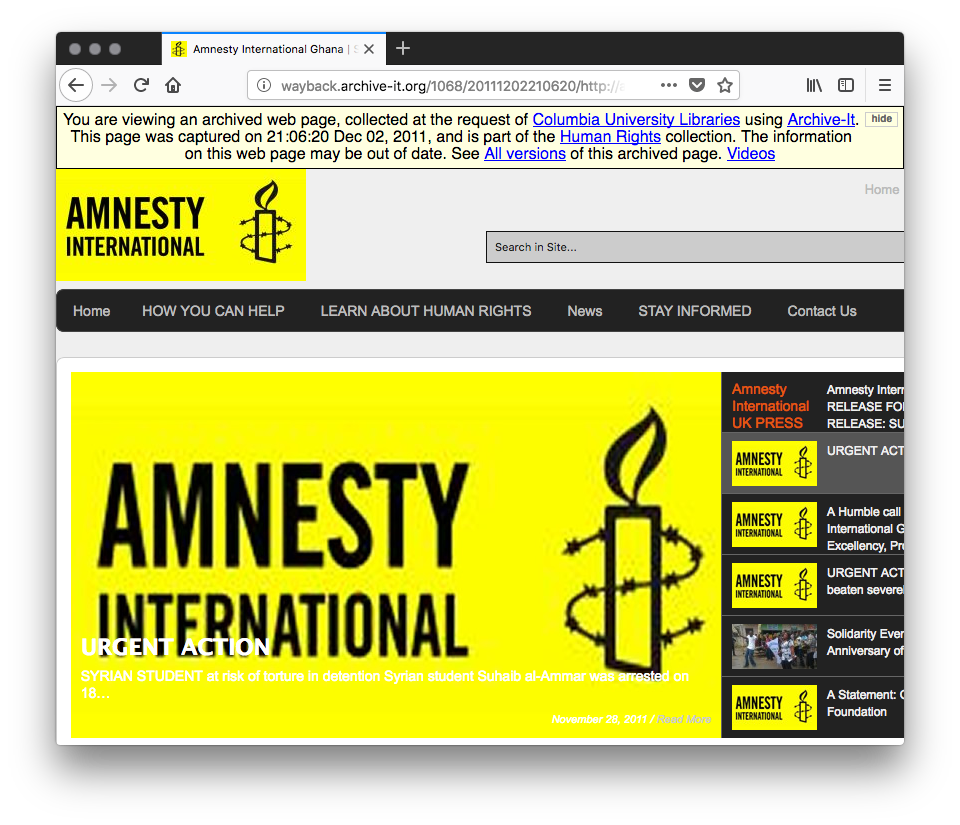}
\includegraphics[width=0.4\textwidth]{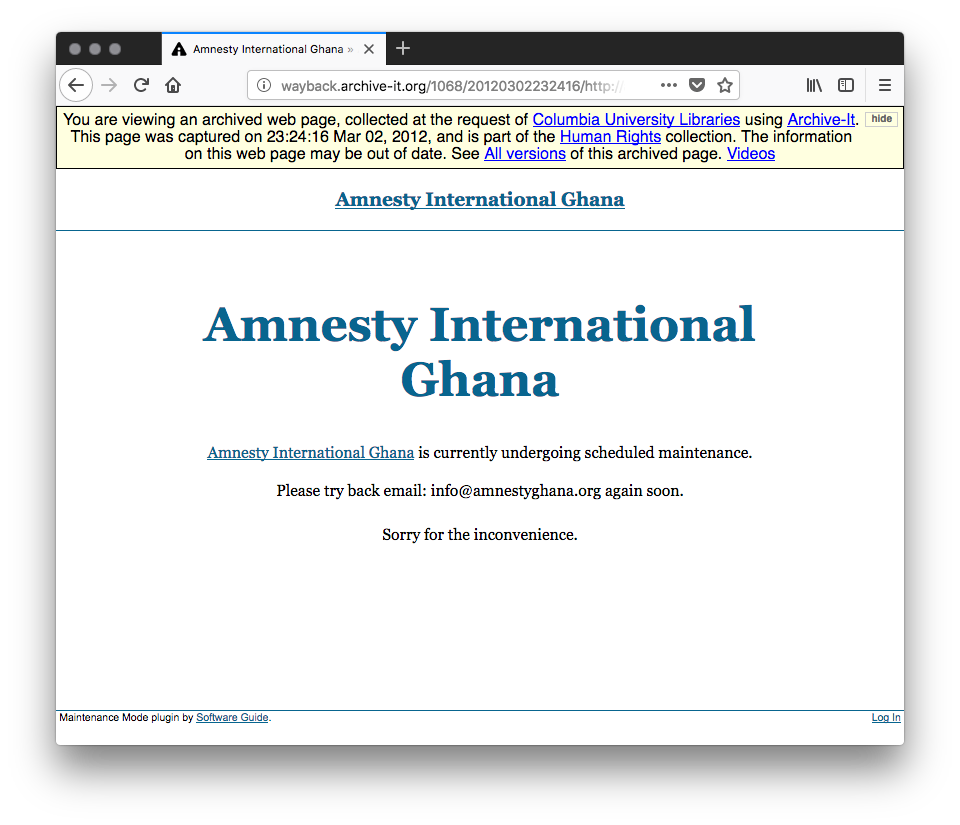}
\caption{The seed \url{http://amnestyghana.org/} preserved in Archive-It's \emph{Human Rights} collection (left) goes off topic due to site maintenance}
\label{fig:maintenance}
\end{figure*}

\begin{figure*}[t]
\includegraphics[width=0.4\textwidth]{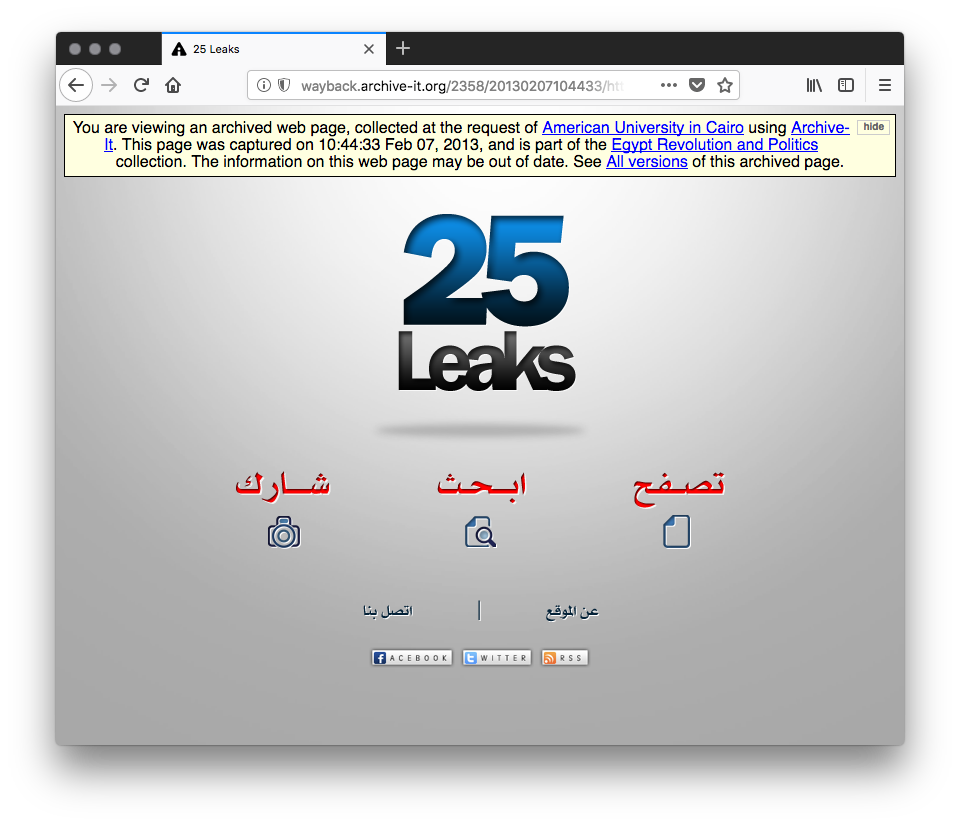}
\includegraphics[width=0.4\textwidth]{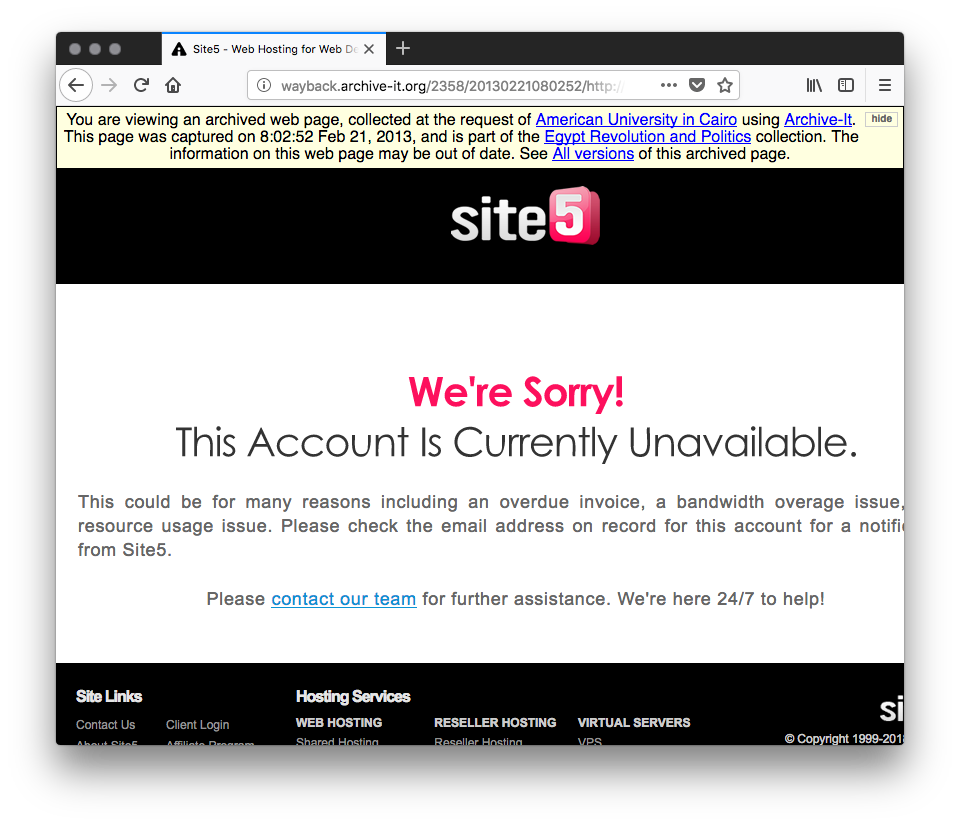}
\caption{The seed \url{http://25leaks.com} preserved in Archive-It's \emph{Egypt Revolution and Politics} collection (left) is later supended due to non-payment.}
\label{fig:suspended}
\end{figure*}

\begin{figure*}[t]
\includegraphics[width=0.4\textwidth]{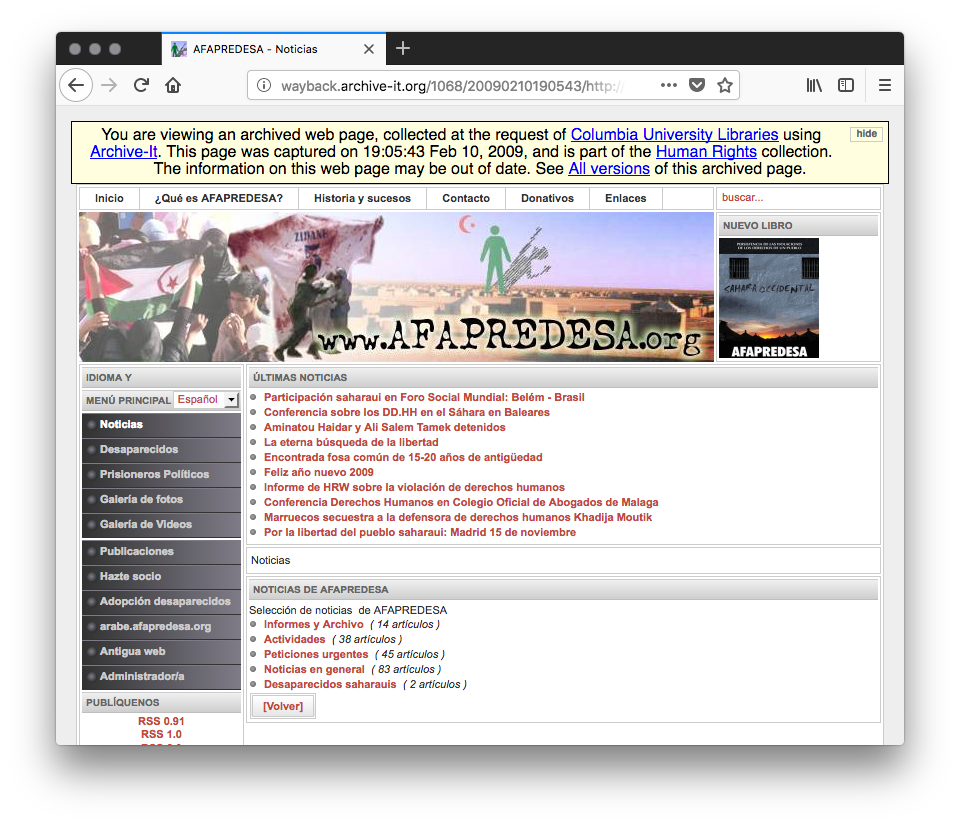}
\includegraphics[width=0.4\textwidth]{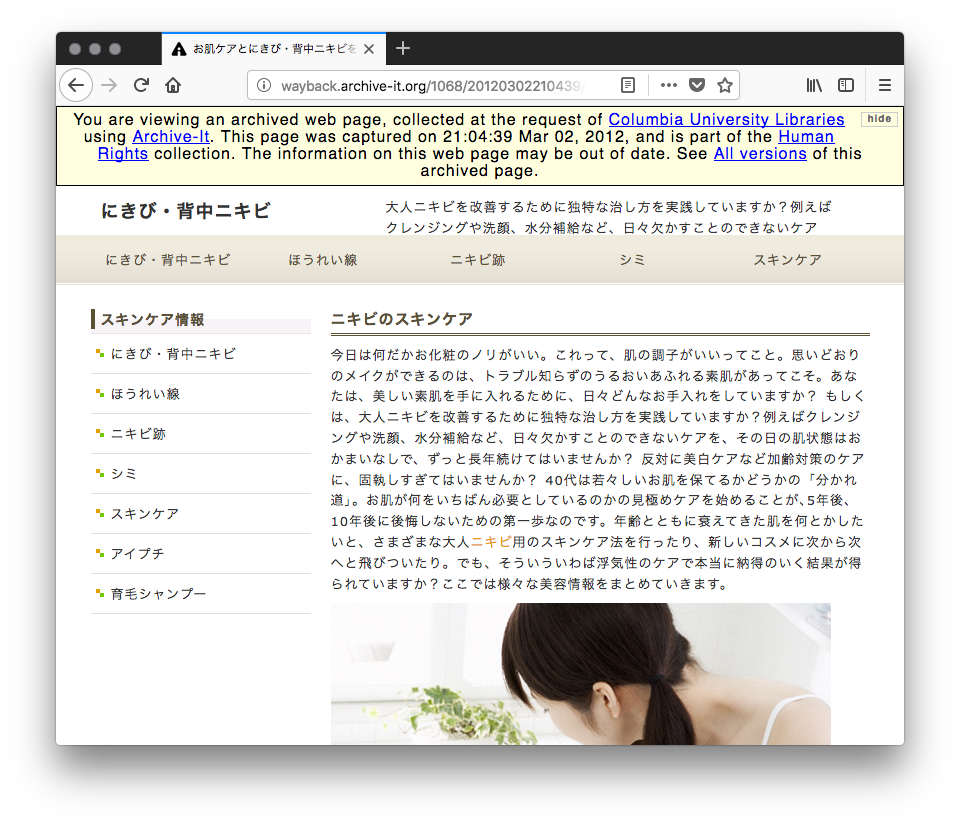}
\caption{The seed \url{http://www.afapredesa.org} preserved in Archive-It's \emph{Human Rights} collection (left) changes ownership to a different organization that publishes in Japanese (right).}
\label{fig:ownership}
\end{figure*}

\begin{figure*}[t]
\includegraphics[width=0.4\textwidth]{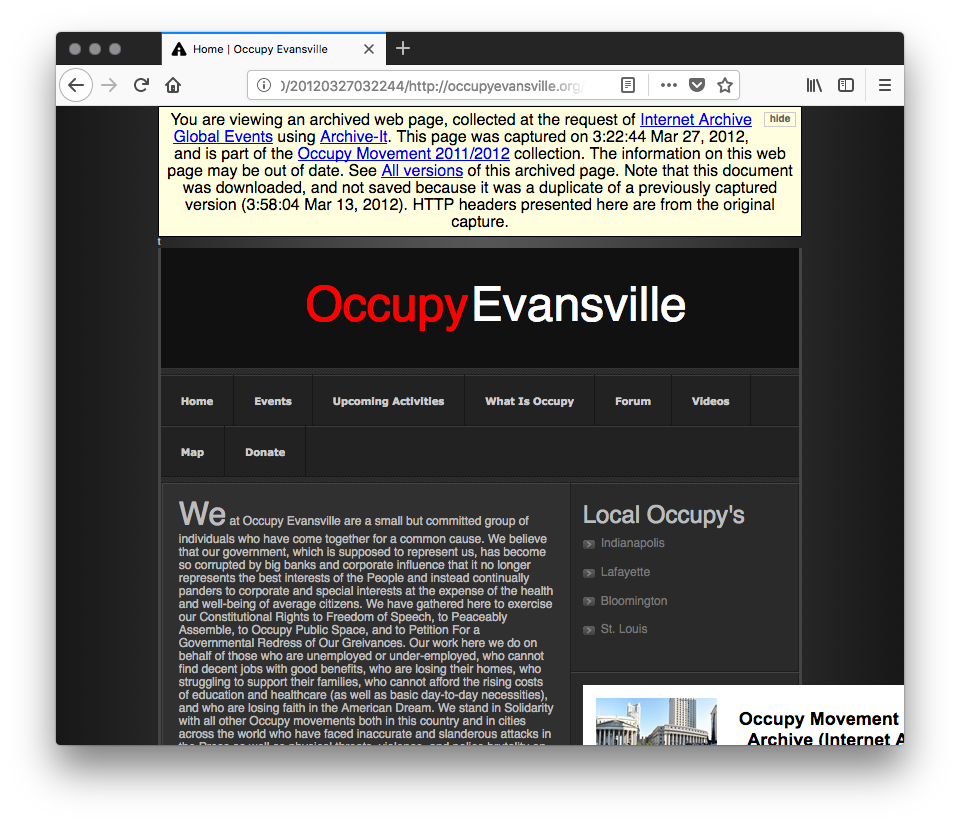}
\includegraphics[width=0.4\textwidth]{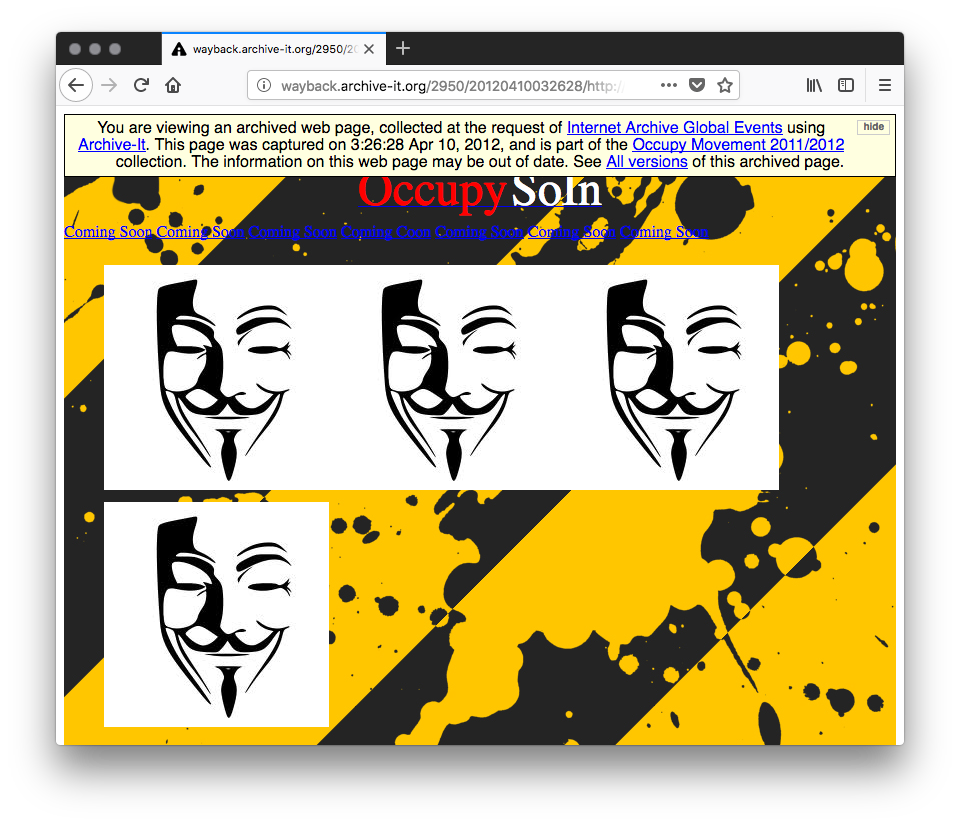}
\caption{The seed \url{http://occupyevansville.org/} preserved in Archive-It's \emph{Occupy Movement 2011/2012} collection (left) was later hacked by Anonymous (right)}
\label{fig:hacked}
\end{figure*}

Consider a collection about the Olympics. The top page of a sports site will be on-topic during the Olympics, but will cover other events once the Olympics have finished. Pages can go off topic for a variety of other reasons and web archives still capture these off topic mementos. Web sites have technical issues (Figure \ref{fig:tech_issues}). Owners take sites down for maintenance (Figure \ref{fig:maintenance}). Hosting services suspend sites (Figure \ref{fig:suspended}). New owners purchase existing domains and replace the  site content (Figure \ref{fig:ownership}). Political regimes change, resulting in news sites changing content \cite{aturban_gaddafi}. Hackers deface pages (Figure \ref{fig:hacked}). Owners restructure sites, resulting in broken links, which are off-topic. Detecting off-topic mementos is important for the development of automated collection summaries \cite{AlNoamany:2017:GSA:3091478.3091508} or finding aids, where off-topic content needs to be detected and excluded, lest it alter the output.  As many as 11\% of the mementos in a collection can be off-topic \cite{AlNoamany2016}. To that end, we have developed the Off-Topic Memento Toolkit (OTMT)\footnote{\url{https://github.com/oduwsdl/off-topic-memento-toolkit}}. It is important to note that OTMT only identifies off-topic mementos, it does not remove them from the collection; indeed, when and how pages went off-topic may be of interest to researchers.


We discuss the different similarity measures available with OTMT version 1.0.0 alpha, how we arrived at the default threshold values for each measure, and mention some of the possible future features for the toolkit. We incorporated several different measures because curators may need to tailor the toolkit's abilities to the content they are evaluating. Our contribution is the review of the effectiveness of different similarity measures for identifying off-topic mementos, especially additional similarity measures not covered in prior work \cite{AlNoamany2016}, and the availability of a software package allowing users to discover off-topic mementos for themselves.

\begin{figure}[t]
\includegraphics[width=0.35\textwidth]{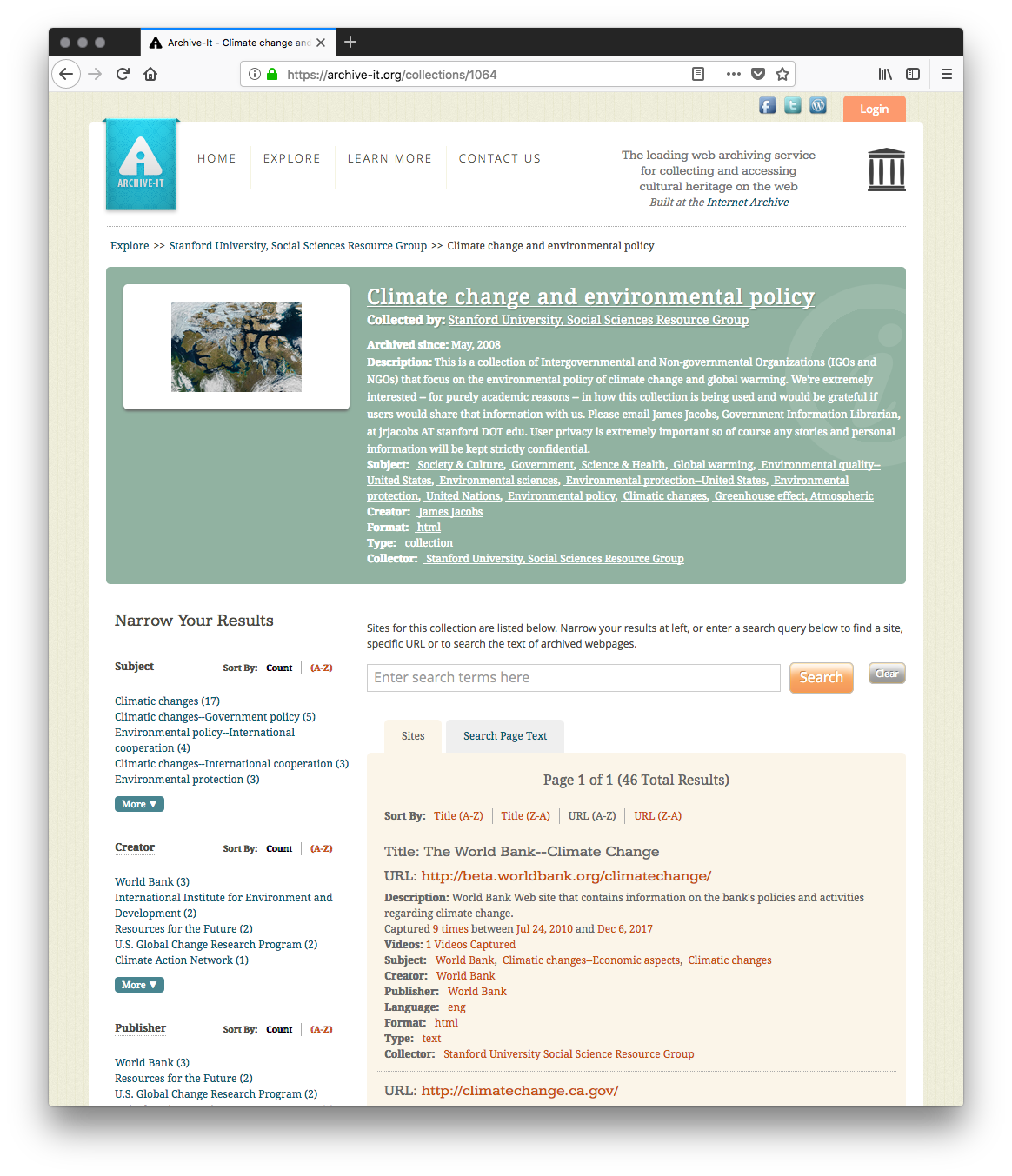}
\caption{The \emph{Climate change and environmental policy} Collection at Archive-It, collected by Stanford University}
\label{fig:archiveit-screenshot}
\end{figure}

\begin{figure}[t]
\begin{lstlisting}[basicstyle=\footnotesize\ttfamily]
<http://bloombergvillenow.org/>; rel="original",
<http://wayback.archive-it.org/2950/timemap/link/http://bloombergvillenow.org/>; rel="self"; type="application/link-format"; from="Tue, 03 Jan 2012 01:43:26 GMT"; until="Thu, 31 May 2012 20:08:41 GMT",
<http://wayback.archive-it.org/2950/http://bloombergvillenow.org/>; rel="timegate",
<http://wayback.archive-it.org/2950/20120103014326/http://bloombergvillenow.org/>; rel="first memento"; datetime="Tue, 03 Jan 2012 01:43:26 GMT",
<http://wayback.archive-it.org/2950/20120109025617/http://bloombergvillenow.org/>; rel="memento"; datetime="Mon, 09 Jan 2012 02:56:17 GMT",
<http://wayback.archive-it.org/2950/20120531200841/http://bloombergvillenow.org/>; rel="last memento"; datetime="Thu, 31 May 2012 20:08:41 GMT"
\end{lstlisting}
\caption{An example TimeMap for the seed URI \url{http://bloombergvillenow.org/} in the Archive-It collection \emph{Occupy Movement 2011/2012}}
\label{fig:timemap-example}
\end{figure}

\section{Background and Related Work}

Web archive collections, like those at Archive-It, are often created along a specific theme. Figure \ref{fig:archiveit-screenshot} displays a screenshot of the \emph{Climate change and environmental policy} collection at Archive-It, collected by the Stanford University Social Sciences Research Group. The curators who built this collection intended for its mementos to contain information on this topic, and thus selected specific seeds to include in this collection. Each seed has a corresponding \textbf{TimeMap} which provides a record of all mementos for that seed \cite{rfc7089}. Figure \ref{fig:timemap-example} displays a TimeMap for the seed URI \url{http://bloombergvillenow.org/}. Each entry with a relation of \texttt{memento} indicates that the URI on that line is a memento for this seed, and its corresponding \texttt{datetime} keyword indicates when this memento was captured, its  \textbf{memento-datetime}. As noted above, seeds sometimes go off-topic, and anyone studying climate change will not want these off-topic mementos in the data that they review.

We will use the Memento terminology in the rest of this paper. TimeMap URIs are abbreviated as \textbf{URI-T} and memento URIs are abbreviated as \textbf{URI-M}.

AlSum \cite{10.1007/978-3-319-06028-6_25} explored techniques for producing a thumbnail for each memento in a TimeMap. Each thumbnail is a resized image produced by taking a screenshot of a memento in a browser. Using AlSum's thumbnails, a curator can review all images visually and mark those that are off-topic. There are Archive-It collections, like the \emph{Goverment of Canada Publications} with 314,032 mementos, that make manual review of each memento a costly endeavor both in terms of time and personnel.

There are many methods of comparing the similarity of two documents. Seeking ways to improve web crawling, Manku \cite{Manku:2007:DNW:1242572.1242592} determined that Simhash \cite{Charikar:2002:SET:509907.509965} is effective at discovering documents that are similar. Adar \cite{Adar:2009:WCE:1498759.1498837} employed the S{\o}rensen-Dice coefficient \cite{sorensen, doi:10.2307/1932409} to understand the changes in content of the same resource over the course of a crawl.  Sivakumar \cite{Sivakumar2015} tested the viability of the Jaccard Index \cite{NPH:NPH37} for improving search results by identifying duplicate advertisements and headers. Hajishirzi \cite{Hajishirzi:2010:AND:1835449.1835520} applied cosine similarity \cite{cosine} to the problem of identifying duplicate news articles. Zittrain \cite{zittrain_albert_lessig_2014} and Jones \cite{10.1371/journal.pone.0167475} have used these methods to analyze content drift in web archive collections. We adopt these \textbf{similarity measures} as a way to determine if a memento is off-topic.

Topic modeling techniques, such as Latent Dirichlet Allocation \cite{Blei:2003:LDA:944919.944937}, typically break a corpus into clusters of documents. Each cluster contains documents that share some topic. We are looking for off-topic documents. Though it is conceivable that the smallest cluster may contain off-topic documents, we have not evaluated that topic modeling techniques behave this way and only consider cosine similarity informed by Latent Semantic Indexing \cite{ASI:ASI1}, a process developed by Radim {\v R}eh{\r u}{\v r}ek \cite{rehurek-dissertation}.

The mementos usually viewed by users of web archives have been augmented for usability and legal reasons, including banners to identify the containing archive. The extra content in these augmented mementos leaves them unsuitable for comparison. Fortunately, Archive-It provides access to \textbf{raw mementos} at special URIs \cite{raw_mementos, raw_mementos_take_two}. These raw mementos contain the original content that was observed by the web archive at the time of capture, without any rewriting. It is these raw mementos that the OTMT uses in its analysis.

Once the OTMT has the content of a raw memento, it still must preprocess it before comparing it using a given similarity measure. Most mementos consist of HTML, JavaScript, and CSS. This boilerplate provides no useful information for the decision as to whether a memento is on or off-topic, and must be removed \cite{pomikalek-justext}. The OTMT then tokenizes, stems, and removes stop words \cite{croft2015} from the resulting text. We refer to these steps as \textbf{preprocessing}.



In 2015, AlNoamany performed a study to detect off-topic mementos \cite{AlNoamany2016}. In that work, she reviewed the patterns that emerge when a collection goes off-topic. Pages can be always on topic or always off topic. Pages can start on topic and then drift off topic permanently at some point. This is usually the case when a page goes offline, falls under new ownership, or the content has drifted far from the original topic. Pages can also oscillate on or off-topic, often due to hacking or technical problems. She sampled mementos from three Archive-It collections and manually labeled these mementos as ``on-topic'' or ``off-topic''. Using this dataset, she then evaluated the effectiveness of different similarity measures. This work was later used to remove off-topic mementos in consideration of generating summaries of Archive-It collections \cite{AlNoamany:2017:GSA:3091478.3091508}. We build upon her work by evaluating additional similarity measures against this same dataset, which we refer to as the \textbf{gold standard dataset}, and have also developed the OTMT for curators to use in discovering off-topic mementos.

\section{TimeMap Measures}

The OTMT uses the TimeMap of each seed to group mementos for comparison. The OTMT supports different similarity measures against the first (i.e., earliest) memento in a TimeMap. The assumption is that the first memento was on-topic when its seed was submitted to the archive and that automated crawling continued afterward at various intervals.


\begin{algorithm}[t]
\begin{algorithmic}[1]
\For{$timemap \in collection$}

	$f \leftarrow HTTP_{GET_{raw}}(memento_{first})$
	
	$f \leftarrow preprocess(f)$
	
	\For{$memento \in timemap$}
		
		$m \leftarrow HTTP_{GET_{raw}}(memento)$

		$m \leftarrow preprocess(m)$

		$score \leftarrow computeSimilarity(f, m, measure)$
		
		$saveScore(score, memento, measure)$
		
	\EndFor
	
\EndFor
\end{algorithmic}
\caption{General algorithm used for all TimeMap measures}
\label{alg:timemap_measures}
\end{algorithm}

The general timemap measure algorithm used by the OTMT is shown in Algorithm \ref{alg:timemap_measures}. This algorithm iterates through all TimeMaps in the collection, dereferencing the raw version of the \textbf{first memento} (denoted by $f$) in the TimeMap for its content. After preprocessing (if necessary), the algorithm iterates through the preprocessed version of every memento in the TimeMap, comparing the first memento to each additional \textbf{considered memento} (denoted by $m$) with the selected similarity measure (denoted by $measure$).

The following sections provide more detail on these measures. As in Algorithm \ref{alg:timemap_measures}, the symbols $f$ and $m$ used in the following equations correspond to the first and considered memento.



\subsection{Structural Measures}

The OTMT provides two structural measures that execute much faster than the others. \textbf{Byte count} tallies the bytes within a memento's content and compares the first memento's bytes with the considered memento. Before calculating the score, only the content of each memento is dereferenced. No preprocessing is performed. Instead of bytes, \textbf{word count} tallies the number of words within a memento's content and compares the number of words in the first memento with the considered memento. Before calculating the score, preprocessing is performed on the memento so that individual words can be counted. The score for each of these measures is based on the percentage difference between the size of the first memento and the considered memento, shown by Equation \ref{eq:wordcount},

\begin{equation}
d_c(f, m) = \left\{
	\begin{array}{ll}
		\frac{c(m) - c(f)}{c(f)} = \frac{c(m)}{c(f)} - 1 & \text{if $m < f$} \\
		0 & \text{if $m \geq f$}
	\end{array}
\right.
\label{eq:wordcount}
\end{equation}

where $c(x)$ indicates the count (byte or word) of memento $x$.

The scores range from $0.0$, meaning the two documents are the same size or larger, to $-1.0$ meaning that the considered memento has reached a size of $0$. We assume that adding content is common if a memento stays on topic. For this reason, we only consider scores that are negative because off-topic pages often only contain short sentences indicating a 404 message, that the web site has failed to pay its bills, or that there is a technical problem. 

\subsection{Set Operation Measures}

Within the OTMT we provide set operations to evaluate the sets of words that make up each document. Because we are interested in the words of each document, the documents are preprocessed before using these measures.

The Jaccard Index (sometimes called Jaccard Coefficient) compares two sets \cite{NPH:NPH37}. To normalize this score, OTMT uses the \textbf{Jaccard distance} as defined by the Python distance library \cite{python-distance} to compare the two sets of words making up the documents. Jaccard distance calculates the percentage of overlap between the words in both documents, as shown in Equation \ref{eq:jaccard},

\begin{equation}
d_J (f, m) = \frac{ \lvert t(f) \cup t(m) \rvert   - \lvert t(f) \cap t(m) \rvert }{ \lvert t(f) \cup t(m) \rvert }
\label{eq:jaccard}
\end{equation}

where $t(x)$ indicates the tokens produced by the preprocessing of the content of memento $x$.



The S{\o}rensen-Dice Coefficient is another method of comparing two sets \cite{sorensen, doi:10.2307/1932409}. The OTMT uses the \textbf{S{\o}rensen-Dice distance} as defined by the Python distance library to compare the two sets of words making up the documents. S{\o}rensen-Dice is different in that it takes twice the number of words in common and divides them by the number of total words, shown in Equation \ref{eq:sorensendistance}.

\begin{equation}
d_S (f, m) =  1 - \frac{ 2 \lvert t(f) \cap t(m) \rvert }{ \lvert t(f) \rvert +  \lvert t(m) \rvert }
\label{eq:sorensendistance}
\end{equation}


Both distance measures have scores ranging from 0.0, meaning that the documents are the same, to 1.0, meaning that the documents are completely dissimilar. Both are different from word count because the individual words in each document are considered.




%
%

\subsection{Simhash Measures}


The OTMT provides Simhash as implemented by the Python Simhash library \cite{python-simhash}. That library's functions allow for multiple types of input: term frequencies or raw content. 

For the \textbf{Simhash of term frequencies} the term frequencies (TF) are provided as input to a corresponding cryptographic hash function, thus preprocessing is needed. The hash of each term frequency then makes up part of a larger hash representing the document.

If the \textbf{Simash of raw content} is desired, then the raw memento content is supplied to the function, and no preprocessing is needed. The function converts the raw document into 4-grams, strings of 4 characters long. A hash is computed on each 4-gram and these hashes make up the resulting Simhash. This form of Simhash is influenced by the position of each 4-gram in the document.

Differences between these smaller hashes are reflected in the resulting Simhash, allowing one to compare two documents by comparing their Simhash values. Simhash scores are calculated based on the number of bits different between the two Simhashes. A score of $0$ indicates that the two strings of bits are the same. A score of $64$ indicates that the two strings are completely different. Because of this, it is highly unlikely that all 64 bits of a Simhash will be different.

\subsection{Cosine Similarity Measures}

Cosine similarity compares two documents as the distance of two vectors.  These vectors can be constructed different ways, but the source mementos always require preprocessing.

The first vector construction method supported by the OTMT is \textbf{TF-IDF}. With this method each document is converted into a vector that represents each word and its term frequency. The values of the vector are further weighted with the inverse document frequency (IDF) \cite{doi:10.1108/eb026526} of every term of the other mementos in the TimeMap. This way, each document vector is informed not by just the first and considered memento, but also all other mementos in the TimeMap. The OTMT uses the TF-IDF functionality of the scikit-learn library \cite{scikit-learn}.

The second vector construction method uses the Latent Semantic Indexing (\textbf{LSI}) \cite{ASI:ASI1} capability of the gensim library \cite{rehurek_lrec}. The OTMT implementation for computing these vectors follows the  gensim Similarity Queries tutorial \cite{gensim-similarities-tutorial}. In this case, the vector of each document is informed by LSI. 




The cosine of the resulting angle of these vectors (either via TF-IDF or LSI) is then used to generate a distance score. Equation \ref{eq:cosine} shows how these vectors produce a score,

\begin{equation}
d_c (f, m) = \frac{ v(f) \cdot v(m) }{  \lVert v(f) \rVert \lVert v(f) \rVert }
\label{eq:cosine}
\end{equation}

where $v(x)$ indicates the vector produced by the content of memento $x$. Cosine similarity ranges from $1$, most similar, to $0$ indicating that the vectors are completely different.

\begin{table}[t]
\centering
\fontsize{7}{8}\selectfont
\caption{Similarity measures supported by the OTMT}
\label{tab:preprocessing}
\begin{tabular}{l | l | l | l | l}
\textbf{Measure}  & \textbf{Fully} & \textbf{Fully} & \textbf{Preprocessing}  & \textbf{OTMT \texttt{-tm}}  \\
& \textbf{Equivalent} & \textbf{Dissimilar} & \textbf{Performed}   & \textbf{keyword} \\
& \textbf{Score} & \textbf{Score} &  \\
\hline
\hline
Byte         &            0.0         & -1.0                & No  & \texttt{bytecount} \\
Count & & & & \\
\hline
Word        &            0.0         & -1.0                & Yes & \texttt{wordcount} \\
Count & & & & \\
\hline
Jaccard  &            0.0         & 1.0                & Yes & \texttt{jaccard} \\
Distance & & & & \\
\hline
S{\o}rensen-Dice     & 0.0         & 1.0              & Yes & \texttt{sorensen} \\
Distance & & & & \\
\hline
Simhash of      & 0   & 64             & Yes & \texttt{simhash-tf} \\
Term & & & & \\
Frequencies & & & & \\
\hline
Simhash of          & 0    & 64          & No & \texttt{simhash-raw} \\
Raw Memento  & & & & \\
Content  & & & & \\
\hline
Cosine     & 1.0   & 0.0           & Yes & \texttt{cosine} \\
Similarity & & & & \\
 of TF-IDF  & & & & \\
 Vectors  & & & & \\
\hline
Cosine  & 1.0 & 0.0 & Yes & \texttt{gensim\_lsi} \\
Similarity & & & & \\
 of LSI  & & & & \\
  Vectors & & & &
\end{tabular}
\end{table}

\section{Toolkit Usage}

The OTMT allows a user to select an input type, one or more similarity measures, and an output file. These arguments indicate the input-measure-output architecture of the OTMT which attempts to separate the concerns of acquiring mementos for comparison (input), measuring those mementos (measure), and then producing results (output). This architecture facilitates the addition of future input types, measures, and outputs.

The toolkit is run from the command line. For example, to evaluate Archive-It collection 7877 using both measures Jaccard distance and bytecount and then save the output to outputfile.json, one would run the command:

\begin{lstlisting}
detect_off_topic -i archiveit=7877 -o outputfile.json -tm jaccard=0.80,bytecount=-0.50
\end{lstlisting}

where \texttt{-i} indicates the input type followed by its arguments, and the \texttt{-o} indicates the name of the output file, and \texttt{-tm} (for \emph{TimeMap Measure}) indicates that the next argument is a list of measures and thresholds. 
Table \ref{tab:preprocessing} shows the available similarity measures, their score ranges, whether preprocessing is performed, and the OTMT keyword used to specify the measure on the command line.

The toolkit supports the following forms of input:
\begin{itemize}
\item One or more Memento TimeMaps (input type \texttt{timemap} followed by a \texttt{=} and then the URIs of TimeMaps separated by commas)
\item One or more WARC \cite{ISO28500} files (input type \texttt{warc} followed by a \texttt{=} and then the filenames of the WARCs separated by commas)
\item An Archive-It collection ID (input type \texttt{archiveit} followed by a \texttt{=} and then the Archive-It collection ID)
\end{itemize}

If an Archive-It collection ID is supplied, then the OTMT extracts all seeds from that collection's Archive-It pages and constructs URI-Ts to discover all mementos for those seeds. If a WARC is supplied, then its contents are extracted and TimeMaps are generated for each original resource.

The default output is in JSON format, as shown in Figure \ref{fig:default_output}. Each URI-T key contains a dictionary of URI-M keys. Each URI-M key contains a dictionary of similarity measures run against that URI-M. For each similarity measure the output indicates which preprocessing was performed on the memento. Each measure record indicates the topic status (on or off-topic) based on the threshold supplied. For each URI-M an overall topic status is listed, which is determined based on whether or not one of the measures determined that the memento was off-topic. The toolkit also supports a CSV version of this same data.

\begin{figure}
\begin{lstlisting}[basicstyle=\footnotesize\ttfamily]
"http://wayback.archive-it.org/1068/timemap/link/http://www.badil.org/": {
        "http://wayback.archive-it.org/1068/20130307084848/http://www.badil.org/": {
            "timemap measures": {
                "cosine": {
                    "stemmed": true,
                    "tokenized": true,
                    "removed boilerplate": true,
                    "comparison score": 0.10969941307631487,
                    "topic status": "off-topic"
                },
                "bytecount": {
                    "stemmed": false,
                    "tokenized": false,
                    "removed boilerplate": false,
                    "comparison score": 0.15971409055425445,
                    "topic status": "on-topic"
                }
            },
            "overall topic status": "off-topic"
        },
        ...
\end{lstlisting}
\caption{Example JSON output record from the OTMT for the memento at URI-M \url{http://wayback.archive-it.org/1068/20130307084848/http://www.badil.org/} from the TimeMap at URI-T \url{http://wayback.archive-it.org/1068/timemap/link/http://www.badil.org/} }
\label{fig:default_output}
\end{figure}

If a threshold value is not provided on the command line for a given measure, then the OTMT uses a reasonable default. In the next section, we detail how we arrived at reasonable default values for each measure.

\section{Evaluation of Reasonable Default Thresholds}

To acquire reasonable default thresholds for each measure, we used the dataset from AlNoamany's work \cite{AlNoamany2016}. Information about the gold standard dataset used in this study is shown in Table \ref{tab:gold-standard-distribution} and the version used in this study is available on GitHub\footnote{\url{https://github.com/oduwsdl/offtopic-goldstandard-data/tree/5139aca762e1ddac76da628436dbc48ae38807f2}}.


\begin{table}[t]
\centering
\small
\caption{Definitions of the conditions used to calculate the $F_1$ score for threshold determinations}
\label{tab:threshold-determination-conditions}
\begin{tabular}{lll}
\textbf{Gold Standard Data} & \textbf{OTMT} & \textbf{Condition} \\
\hline
On-Topic                    & On-Topic      & True Negative      \\
\hline
Off-Topic                   & On-Topic      & False Negative     \\
\hline
On-Topic                    & Off-Topic     & False Positive     \\
\hline
Off-Topic                   & Off-Topic     & True Positive     
\end{tabular}
\end{table}

\begin{table}[t]
\centering
\small
\caption{Distribution of the Gold Standard Data Set}
\label{tab:gold-standard-distribution}
\begin{tabular}{llrrr}
\textbf{Collection} & \textbf{Collection}               & \textbf{\# seeds} & \textbf{\# mementos} & \textbf{\# off-topic} \\
\textbf{ID} & \textbf{Name} & \textbf{in sample} & \textbf{in sample} & \textbf{} \\
\hline
1068                 & Human                  & 199                & 2302                  & 95 (4\%)           \\
& Rights & & & \\
\hline
2358                 & Egypt  & 136                & 6886                  & 384 (6\%)          \\
& Revolution & & & \\
& and Politics & &  & \\
\hline
2950                 & Occupy      & 255                & 6569                  & 458 (7\%) \\
& Movement & & & \\
& 2011/2012 & & & \\
\end{tabular}
\end{table}


For evaluating the OTMT, we focused on the $F_1$ measure,  seen in Equation \ref{eq:f1_measure},

\begin{equation}
F_1 = \frac{2TP}{2TP + FP + FN}
\label{eq:f1_measure}
\end{equation}

where $TP$ indicates the number of true positives, $FP$ indicates the number of false positives, and $FN$ indicates the number of false negatives. Table \ref{tab:threshold-determination-conditions} shows the conditions we used to calculate these values for each measure and threshold combination. 

For comparison with AlNoamany's results, we provide values for a second metric of \textbf{accuracy}, shown in Equation \ref{eq:acc_measure} using the same symbols as Equation \ref{eq:f1_measure}, where $TN$ indicates the number of true negatives and the other symbols have the same meaning as Equation \ref{eq:f1_measure}.

\begin{equation}
Accuracy = \frac{(TP + TN)}{(TP + FP + FN + TN)}
\label{eq:acc_measure}
\end{equation}



\begin{figure*}[t]

\begin{subfigure}{0.33\textwidth}
\includegraphics[width=1\textwidth]{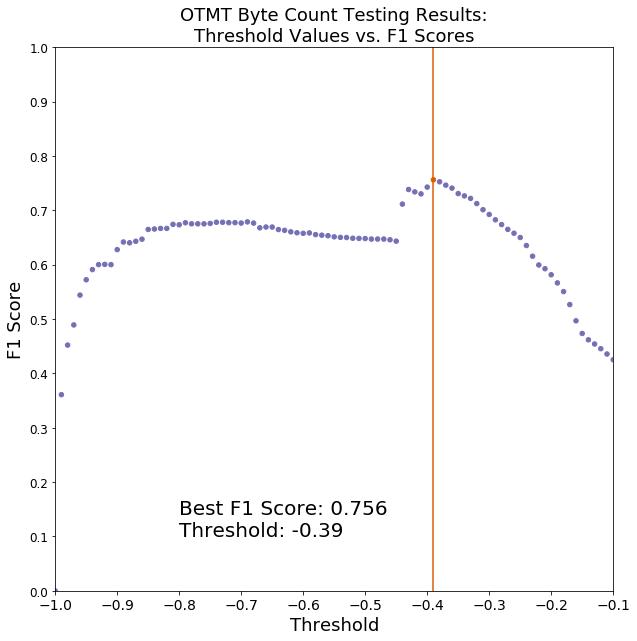}
\caption{Byte Count}
\label{fig:bytecount-threshold}
\end{subfigure}%
~
\begin{subfigure}{0.33\textwidth}
\includegraphics[width=1\textwidth]{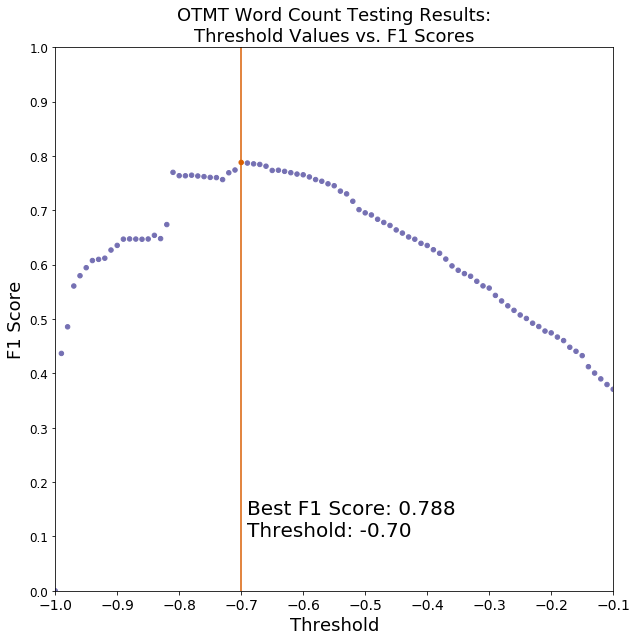}
\caption{Word Count}
\label{fig:wordcount-threshold}
\end{subfigure}%
~
\begin{subfigure}{0.33\textwidth}
\includegraphics[width=1\textwidth]{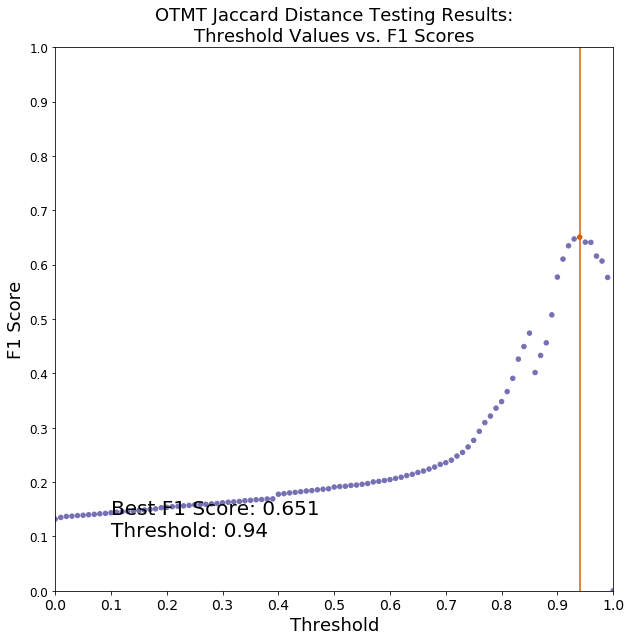}
\caption{Jaccard Distance}
\label{fig:jaccard-threshold}
\end{subfigure}

\begin{subfigure}{0.33\textwidth}
\includegraphics[width=1\textwidth]{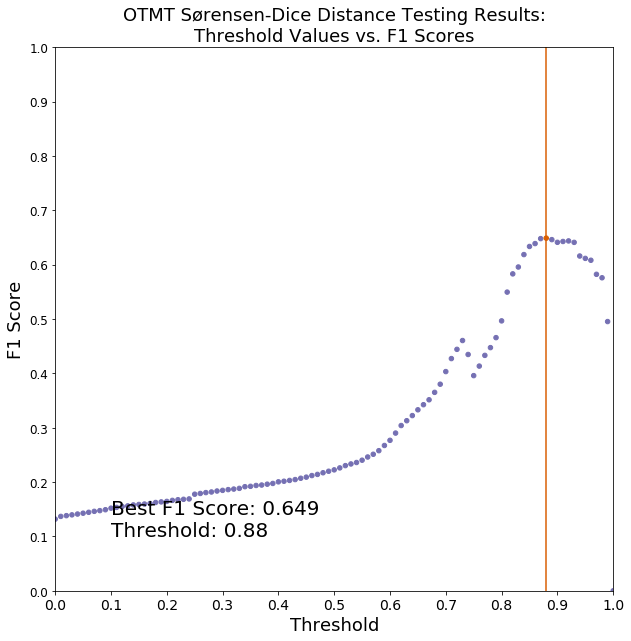}
\caption{S{\o}rensen-Dice Distance}
\label{fig:sorensen-threshold}
\end{subfigure}%
~
\begin{subfigure}{0.33\textwidth}
\includegraphics[width=1\textwidth]{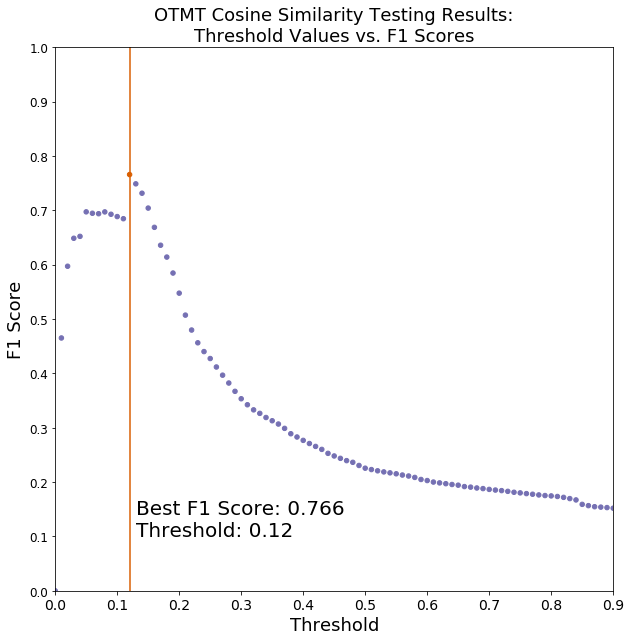}
\caption{Cosine Similarity of TF-IDF Vectors}
\label{fig:cosine-threshold}
\end{subfigure}

\begin{subfigure}{0.33\textwidth}
\includegraphics[width=1\textwidth]{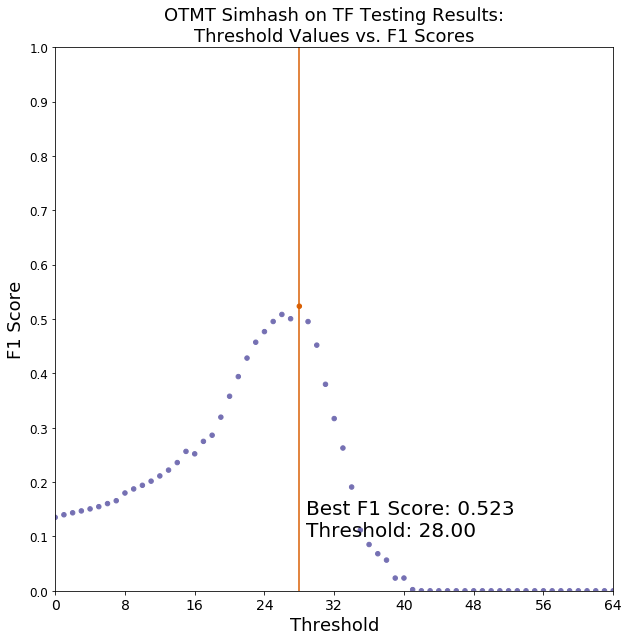}
\caption{Simhash of Term Frequencies}
\label{fig:tf_simhash-threshold}
\end{subfigure}%
~
\begin{subfigure}{0.33\textwidth}
\includegraphics[width=1\textwidth]{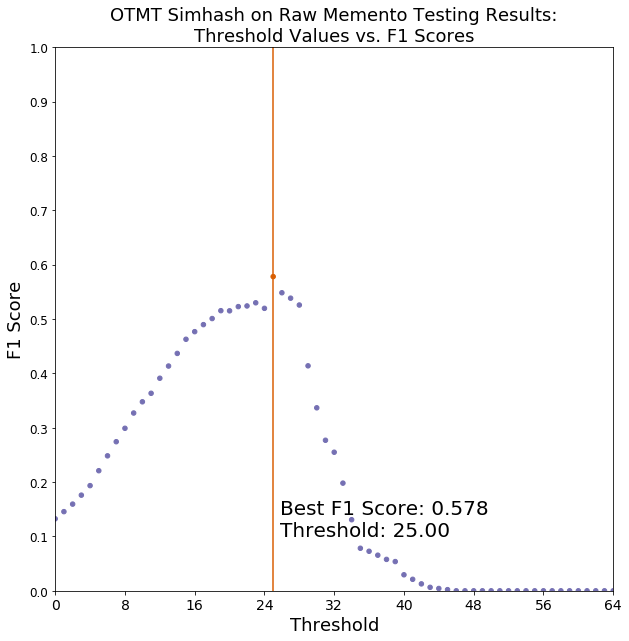}
\caption{Simhash of raw memento content}
\label{fig:raw_simhash-threshold}
\end{subfigure}%
~
\begin{subfigure}{0.33\textwidth}
\includegraphics[width=1\textwidth]{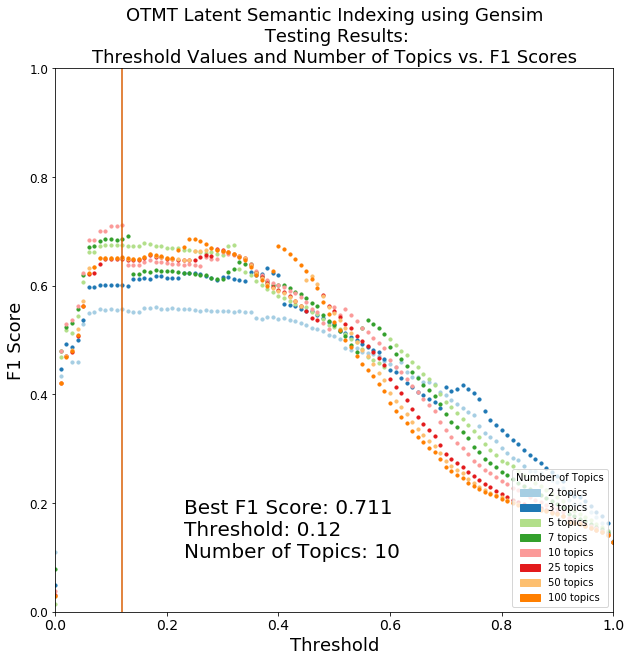}
\caption{Cosine Similarity of LSI Vectors (different colors represent different numbers of topics)}
\label{fig:gensim_lsi-threshold}
\end{subfigure}

\caption{Scatter plots of threshold $F_1$ testing results for different similarity measures}
\label{fig:scatter_thresholds}

\end{figure*}

AlNoamany tested each similarity measure with 21 thresholds \cite{AlNoamany2016}. To get more precise threshold values for each measure we ran the OTMT and then saved the resulting scores from the comparison of each first memento and considered memento. We then iterated through each memento in the output. Starting with a lower limit as the threshold, we tested each score against that limit and declared that memento as on or off-topic depending on the value of the threshold, the score, and the direction of the comparison operator (e.g., \textgreater ~ or \textless) for that measure. We then incremented the threshold and compared again, saving the off-topic determination again. This process was repeated until we reached a designated upper limit. Visualizations of the results for each measure are shown in Figure \ref{fig:scatter_thresholds} with each threshold on the x-axis and the resulting $F_1$ score on the y-axis.


For example, with Byte Count, we ran the OTMT and saved the scores for each memento. We declared a memento off-topic if its score was less than -0.99. We then took each memento's byte count score and declared it off topic if its score was less than -0.98. We then repeated for each memento with a threshold set at -0.97. This process was repeated, in increments of 0.01 until we reached 0.

We then compared the off-topic determinations per threshold with the gold standard data. From there, we were able to generate a corresponding $F_1$ score for each threshold value.

We had assumed that testing the thresholds in this way would help us discover threshold values close to those found by AlNoamany, and we did get close in some cases. The $F_1$ scores, however, are often worse. The OTMT uses the justext library \cite{pomikalek-justext} for boilerplate removal, whereas AlNoamany used boilerpipe \cite{Kohlschutter:2010:BDU:1718487.1718542}. OTMT uses nltk \cite{nltk-book} for tokenization and stemming whereas AlNoamany used scikit-learn \cite{scikit-learn}. These subtle differences in libraries combined with gold standard data set updates, download errors, changes in how Archive-It handles mementos now compared with 2015, and differences in preprocessing techniques, are likely the reason for these differences.

Byte count threshold scores between -1 and 0 were tried, at increments of 0.01. A threshold score of -0.39 produces the best $F_1$ score. This means that the off-topic memento is 39\% smaller than the first memento in the TimeMap. AlNoamany's findings suggested a threshold value of -0.65, making the OTMT results more strict. This is one case where our $F_1$ score of 0.756 is higher than AlNoamany's finding of 0.584.

We used the same range and increments to test word count. A threshold score of -0.70 produces the best $F_1$ score. This means that the off-topic memento has 70\% fewer words than the first memento in the TimeMap. AlNoamany's findings suggested a threshold value of -0.85, again making the OTMT results of -0.70 more strict.

For Jaccard and S{\o}rensen-Dice we used the same range of threshold values from $0$ to $1$, at increments of 0.01. A threshold score of 0.94 has the best $F_1$ value for Jaccard. AlNoamany discovered that a value of 0.05 was best, but her work used the pure Jaccard Index rather than the Jaccard distance. Seeing as the Jaccard distance is $1 -$ Jaccard index, this threshold is consistent with her findings. The $F_1$ score of 0.651, however, is  greater than her $F_1$ result of 0.538 for Jaccard.

S{\o}rensen-Dice was not attempted by AlNoamany. Its best threshold value is close to that of Jaccard at 0.88. In both cases, the document must be quite dissimilar to be marked as off-topic.


Simhash scores range from $0$ to $64$, based on the number of bits different between simhashes. AlNoamany did not test Simhash. For Simhash based on term frequencies, the best $F_1$ score is achieved if one uses a threshold value of 28 bits. If just the raw content is fed into the function, the best $F_1$ score is achieved at 25 bits.

As noted above, cosine similarity scores range from $1$ to $0$. Very close to AlNoamany's threshold result of 0.15 for the cosine of TF-IDF vectors, the highest $F_1$ score is achieved by the OTMT at a threshold of 0.12. This score is very close to the score of complete dissimilarity between documents. AlNoamany achieved an $F_1$ score of 0.881 compared to our 0.766.

The cosine similarity of LSI vectors has an $F_1$ score of 0.711. The LSI algorithm requires that one specify the number of topics into which one should break up the corpus. We tried values of 2, 3, 5, 7, 10, 25, 50, and 100 topics. The value of 10 worked best for testing with our gold standard data. Unfortunately, there is an element of randomness in the results produced by LSI in gensim \cite{random_lsi}. Five different runs with LSI produced $F_1$ scores near or at 0.711, but their corresponding thresholds ranged from 0.08 to 0.12. We took the mean of the scores and set the default threshold at 0.10.

The best test scores for each measure are shown in Table \ref{tab:otmt-toolkit-measures-sorted}. AlNoamany's results are shown for comparison.  Word Count has the best $F_1$ score, followed by Cosine Similairty of TF-IDF Vectors. Byte Count and Cosine Similarity of LSI Vectors score at third and fourth place, respectively.

\begin{table*}[t]
\centering
\small
\caption{OTMT TimeMap measures sorted by best $F_1$ score and compared with AlNoamany's results}
\begin{threeparttable}
\begin{tabular}{l||r||r|r||r|r|r}
\textbf{}                               & \multicolumn{3}{c||}{\textbf{AlNoamany's results}}                                                                                   & \multicolumn{3}{c}{\textbf{Results of this study}}                                             \\
\hline
\textbf{Similarity}             & \textbf{Best $F_1$} & \textbf{Corresponding} & \textbf{Corresponding}                                     & \textbf{Best $F_1$} & \textbf{Corresponding} & \textbf{Corresponding} \\
\textbf{Measure} & \textbf{Score} & \textbf{Accuracy} & \textbf{Threshold} & \textbf{Score} & \textbf{Accuracy} & \textbf{Threshold}\\
\hline
Word Count                              & 0.806                     & 0.982                           & -0.85                                                                & 0.788                     & 0.971                           & -0.70                            \\
\hline
Cosine Similarity of TF-IDF Vectors     & 0.881                     & 0.983                           & 0.15                                                                 & 0.766                     & 0.965                           & 0.12                             \\
\hline
Byte Count                              & 0.584                     & 0.962                           & -0.65                                                                & 0.756                     & 0.965                           & -0.39                            \\
\hline
Cosine Similarity of LSI Vectors & \multicolumn{3}{c||}{Not tested}                                                                                                     & 0.711                     & 0.965                           & 0.10 with 10 topics   $\dagger$           \\
\hline
Jaccard Distance                        & 0.538                     & 0.962                           & 0.95*  & 0.651                     & 0.953                           & 0.94                             \\
\hline
S{\o}rensen-Dice Distance                  & \multicolumn{3}{c||}{Not tested}                                                                                                     & 0.649                     & 0.953                           & 0.88                             \\
\hline
Simhash on raw memento content          & \multicolumn{3}{c||}{Not tested}                                                                                                     & 0.578                     & 0.934                           & 25                               \\
\hline
Simhash on TF                           & \multicolumn{3}{c||}{Not tested}                                                                                                     & 0.523                     & 0.942                           & 28                               \\
\end{tabular}
\begin{tablenotes}
\item[]{* A derived value is shown for easier comparison. AlNoamany's threshold was 0.05, but she used Jaccard Index rather than Distance.}
\item[]{$\dagger$ LSI is non-deterministic. This threshold value is a mean of several runs.}
\end{tablenotes}
\end{threeparttable}
\label{tab:otmt-toolkit-measures-sorted}
\end{table*}

Can we do better by using different measures together? Structural measures require less time to execute than more semantic measures like cosine similarity. If we can short-circuit the process with a structural measure then we can eliminate those off-topic mementos prior to review with a more time-intensive measure.  AlNoamany did so and found that Word Count and Cosine Similarity worked best. The OTMT accepts multiple measures and evaluates their results as a \emph{logical or}. If at least one of the measures scores a memento as off-topic, then that memento is marked as off topic. We tried different combinations of thresholds just as before and recorded the corresponding $F_1$ and accuracy scores. Table \ref{tab:combined-measures} displays these combinations. The cosine of LSI vectors scores a slightly higher $F_1$ and the same accuracy as word count. The word count score appears to exert more influence than its partner measures in all cases where it is present, making both cosine measures require stricter thresholds to be successful. It does not appear that combining these measures improves the $F_1$ score in the OTMT.


\begin{table}[t]
\centering
\small
\caption{The top 4 scoring measures combined in groups of 2}
\label{tab:combined-measures}
\begin{tabular}{llrr}
\textbf{Measure}          & \textbf{Best}      & \textbf{Corresponding}  & \textbf{Corresponding} \\
& \textbf{$F_1$ Score} & \textbf{Thresholds}  & \textbf{Accuracy} \\
\hline
Cosine of LSI,              & 0.789 & (0.01, -0.70) & 0.971                     \\
 Word Count & & & \\
\hline
Cosine of TF-IDF,             & 0.788 & (0, -0.70)  & 0.971                     \\
Word Count & & & \\
\hline
Word Count,                & 0.788 & (-0.70, -0.94)  & 0.971                    \\
Byte Count & & & \\
\hline
Cosine of TF-IDF,           & 0.766 & (0.12, -0.95) & 0.965                     \\
Byte Count & & & \\
\hline
Cosine of LSI,       & 0.766  & (0.12, 0.12)   & 0.965                     \\
Cosine of TF-IDF & & & \\
\hline
Cosine of LSI,              & 0.759 & (0.01, -0.39) & 0.965                    \\
Byte Count & & & 
\end{tabular}
\end{table}

\section {Future Work}

We intend to improve the OTMT over time. For example, we intend to explore making LSI scores reproducible by establishing a specific random number generator seed \cite{random_lsi}. We have also considered additional TimeMap measures like Spamsum \cite{KORNBLUM200691}, which works like Simhash and was used to demonstrate memento content drift by Jackson \cite{jackson2015}. The OTMT also includes an experimental implementation of cosine of Latent Dirichlet Allocation (LDA) \cite{Blei:2003:LDA:944919.944937} vectors from the Gensim library \cite{rehurek_lrec}. This implementation generates errors for some TimeMaps possibly due to a mismatch between the number of topics and the number of features generated by gensim. We have not yet addressed this issue and do not recommend the use of this measure at this time.

Where the previous list of measures were run against all mementos in a TimeMap, it is also conceivable that one can compare each memento in a collection against the collection as a whole. The OTMT does not yet support any measures with this concept. We do envision that such measures could be easily introduced to the OTMT using its existing input-measure-output architecture.

We anticipate the need for curators to have finer grained control over removing boilerplate, stemming, stop word removal, and tokenization. It would also be useful for curators to be able to select different boilerplate removal libraries. We selected justext based on a survey of boilerplate removal methods \cite{boilerplate_survey}, but curators may find that other boilerplate removal libraries work better for their use cases.

\section{Conclusion}

For researchers, identifying off-topic mementos is an important first step to analyzing any web archive collection corpus. We have implemented the Off-Topic Memento Toolkit (OTMT) version 1.0.0 alpha as a way for researchers to identify mementos that are off-topic so that these mementos can be excluded from downstream processing. Because different collections have different needs, we have provided the following similarity measures for use in detecting off-topic mementos: 
\begin{itemize}
\item byte count
\item word count
\item Jaccard distance
\item S{\o}rensen-Dice distance
\item Simhash of term frequencies
\item Simhash of raw memento content
\item cosine similarity of TF-IDF vectors
\item cosine similarity of Latent Semantic Indexing (LSI) vectors
\end{itemize}

Using a gold standard dataset from a prior study, we have evaluated each measure in terms of effectiveness at determining whether a memento is off-topic. We iterated through many threshold values for each measure, and recorded the $F_1$ scores at each threshold. We discovered that word count has the best $F_1$ score, followed by cosine of TF-IDF vectors. Combining the measures did not improve the result, as suggested by prior work.

We present the Off-Topic Memento Toolkit to the world in the hopes that it will help web archive collection curators save time and resources by identifying off-topic mementos.

\begin{acks}
This work has been supported in part by the Institute of Museum and Library Services (LG-71-15-0077-15) and the Andrew Mellon Foundation through the Columbia University Libraries Web Archiving Incentive program.
\end{acks}

\bibliographystyle{ACM-Reference-Format}
\bibliography{bibliography.bib}

\end{document}